\documentclass{article}
\usepackage{slashed}
\usepackage{amsmath}
\usepackage{amssymb}
\usepackage{amsfonts}
\usepackage{color}
\usepackage{graphicx}
\usepackage{subcaption}
\usepackage{url}
\usepackage[margin=1.0in]{geometry}
\newcommand{\nin}{\noindent}
\newcommand{\be}{\begin{equation}}
\newcommand{\ee}{\end{equation}}
\newcommand{\bea}{\begin{eqnarray}}
\newcommand{\eea}{\end{eqnarray}}

\newcommand{\nn}{\nonumber\\}

\graphicspath{{Figures/}} 

\begin{document}

\nin KCL-PH-TH/2018-15

\vspace{1cm}

\begin{center}
{\Large{\bf Black hole interference patterns in flavour oscillations}}

\vspace{0.5cm}

Jean Alexandre${}^{a,}$\footnote{jean.alexandre@kcl.ac.uk}~ and Katy Clough${}^{b,}$\footnote{katy.clough@phys.uni-goettingen.de}
\\[1em] \small{${}^a$ Department of Physics, King's College London,\\ 
\small{London WC2R 2LS, United Kingdom}\\ 
\small{${}^b$  Institut f\"ur Astrophysik, Georg-August-Universit\"at,}\\ 
\small{G\"ottingen, Germany}}

\vspace{1cm}

{\bf Abstract}

\end{center}

Motivated by neutrino astronomy, we consider a plane wave of coupled and massive flavours, scattered by a static black hole, and describe analytically and numerically the corresponding oscillation probability in the surrounding space. Both the interpretation as particles travelling along geodesics and as scattered waves are studied, and consistently show a non-trivial and potentially long range interference pattern, in contrast to the spatially uniform transition probability in a flat spacetime. We introduce a numerical method for studying the oscillations around black holes, which accounts for the full curved geometry and flavour wave mixing. Whilst limited to the region immediately around the black hole, this numerical approach has the potential to be used in more general contexts, revealing the complex interference patterns which defy analytic methods.

\vspace{1cm}

\section{Introduction} 

The emerging field of neutrino astronomy at observatories such as the Ice Cube \cite{TheIceCube:2018gnp}, Antares \cite{Collaboration:2011nsa}, KM3Net 2.0 \cite{Adrian-Martinez:2016fdl} and the Baikal Neutrino Telescope \cite{Avrorin:2011zza} promises to improve our understanding of the fundamental properties of neutrino physics. If these are to be combined with multi messenger observations from gravitational wave events at LIGO/Virgo (\cite{Adrian-Martinez:2016xgn} and see \cite{Sturani:2018axq} for a review), an understanding of the behaviour of neutrinos in curved spacetimes, in particular their flavour oscillations, will be required for proper interpretation of the results. 
Neutrino interferometry has, for example, been proposed as a means by which to discover asymmetry in neutrinos and antineutrinos, and thus distinguish Dirac and Majorana neutrinos \cite{Wright:2017jwl, Gutierrez:2005ax}.

The phenomenon of flavour oscillations is described by Quantum Mechanics and, as explained in more detail below, is based on the study of the wave function of neutrinos, more specifically of their phase.  
The phase shift induced by gravitational fields on particles' wave functions was originally observed with neutrons in the Earth's gravitational field \cite{Colella:1975dq}, with the corresponding effect first studied theoretically in \cite{Stodolsky:1978ks}. For neutrinos oscillations, the effect of gravitational fields on the wave function along geodesics should also have phenomenological implications \cite{Ahluwalia:1996ev}, due for example to the gravitational redshift which increases the oscillation length \cite{Cardall:1996cd}.

General studies of gravitational effects on the phase of wave functions, for different spins, were done in \cite{Alsing:2000ji}, \cite{Zhang:2000mi}, \cite{Visinelli:2014xsa}, \cite{Ramezan:2010zz}, \cite{Sorge:2007zza}, and more specific studies involve supernovae explosions \cite{Yang:2017asl}, neutrino optics and the Lens-Thirring effect \cite{Lambiase:2005gt}, or the study of three flavour oscillations in curved space time \cite{Zhang:2016deq}.
There is also a potential violation of the Equivalence Principle in relation to flavour oscillations in curved space time, as discussed in \cite{Lambiase:2001jr}, \cite{Camacho:1999hv}, \cite{Bhattacharya:1999na}.

We study here the effect of a curved background on the flavour transition probability, motivated by the scenario where neutrinos emitted by an astrophysical event pass by a BH before being observed. Such a scenario has been studied in \cite{Fornengo:1996ef}, where two neutrino beams are lensed by a massive object, and the oscillation probability is calculated at the focal point. We propose instead to assume an incident plane wave of two mixed flavours, which is scattered by a BH. In this stationary process, we calculate analytically and numerically the phase shift which is gravitationally induced at every point of space, and which allows us to determine the oscillation interference pattern. This study therefore takes into account delocalised effects arising from a wave description, and not only the interference of two specific geodesics. 

Section \ref{sec-flavourocillations} reviews features related to oscillation probability, in flat spacetime and also in the presence of a BH. In the latter case, the description is based on particles travelling along geodesics, and we approximate an incident plane wave as a set of parallel and coherent beams, with different impact factors. In this cylindrically symmetric configuration, the resulting scattering pattern exhibits paraboloids of minimum probability, aligned with the axis of symmetry, at large distances. Although highly localised, these paraboloids would in principle result in directions far from the BH along which no oscillation of neutrino flavours would be observed.

We then turn to the wave description in Section \ref{sec-massivewave}, where we give generic analytical features of a massive wave scattered by a BH. The resulting equations can be solved analytically only in very limited cases, and the aim of this section, together with Appendix A, is mainly to understand the scattering properties of the model. 

Section \ref{sec-numrel} covers the numerical implementation of the study, in the vicinity of the BH, following the dynamical evolution of plane wave flavour mass eigenstates on a fixed Schwarzschild background in horizon penetrating coordinates. The resulting interference pattern thus takes into account the self interference of the waves, the full effect of the curved geometry and the resulting backward scattering. Appendix B gives further details of the code used and convergence tests. Whilst we are far from being able to directly observe neutrinos in the vicinity of BHs, such interference patterns may affect the weak interactions around black holes or massive objects such as neutron stars, for example, in a merger scenario.

\section{Flavour oscillations} 
\label{sec-flavourocillations}

We consider here two scalar flavours, since the essence of the flavour oscillation process is independent of the spin.
Also, for fermions, the spin flip effect induced by gravity actually does not occur in Schwarzschild spacetime \cite{Piriz:1996mu}.
The present study is based on plane waves, although this assumption is associated with several ambiguities in the context of flavour oscillations. As explained in the clear and thorough review \cite{Beuthe:2001rc}, these ambiguities can be avoided with wave packets. Nevertheless, the plane wave assumption is enough to construct the essential features of the interference pattern we are interested in. 

\subsection{Oscillation probability in the absence of gravity}\label{absence}

Consider two scalar flavours $\phi_a,\phi_b$, which satisfy the equation of motion 
\be\label{equamot}
\Box\Phi+{\cal M}^2\Phi=0~,
\ee
where
\be
\Phi=\begin{pmatrix} \phi_a \\ \phi_b \end{pmatrix}~~~~\mbox{and}~~~~{\cal M}^2=\begin{pmatrix} m_a^2 & \mu^2 \\ \mu^2 & m_b^2 \end{pmatrix}~.
\ee
The eigenmasses and the corresponding mass eigenstates are given by
\bea
m_\pm^2&=&\frac{1}{2}(m_a^2+m_b^2)\pm\frac{1}{2}\sqrt{(m_a^2-m_b^2)^2+4\mu^4}\\
\begin{pmatrix} \phi_+ \\ \phi_- \end{pmatrix}&=&\begin{pmatrix} \cos\alpha & \sin\alpha \\ -\sin\alpha & \cos\alpha \end{pmatrix}
\begin{pmatrix} \phi_a \\ \phi_b \end{pmatrix}\nonumber
\eea
where $\alpha$ is the mixing angle and
\bea
\sin\alpha&\equiv&\frac{1}{\sqrt2}\sqrt{1-\frac{|m_a^2-m_b^2|}{\Delta m^2}}\\
\Delta m^2&\equiv&m_+^2-m_-^2=\sqrt{(m_a^2-m_b^2)^2+4\mu^4}~.\nonumber
\eea
The equation of motion satisfied by the mass eigenstates is 
\be
\Box\phi_\pm+m_\pm^2\phi_\pm=0~,
\ee
with equal-momentum plane wave solutions in flat space, given by
\be
\phi_\pm=\exp\left(-i\omega_\pm t+i\vec k\cdot\vec r\right)~~~~\mbox{where}~~~~\omega_\pm=\sqrt{m_\pm^2+k^2}~.
\label{eqn:planewave}
\ee
Rotating back to the flavour eigenstates leads to flavour configurations 
\be\label{initial}
\begin{pmatrix} \phi_a \\ \phi_b \end{pmatrix}=\exp\left(i\vec k\cdot\vec r\right)\begin{pmatrix} \cos\alpha & -\sin\alpha \\ \sin\alpha & \cos\alpha \end{pmatrix}
\begin{pmatrix} \exp\left(-i\omega_+t\right) \\ \exp\left(-i\omega_-t\right) \end{pmatrix}~.
\ee
To calculate the oscillation probability, one introduces at every point of space the normalised kets
\bea
\left|\pm\right>&=&e^{i\varphi_\pm}\left|\pm,0\right>\\
\left<+,0\right|\left.+,0\right>&=&\left<-,0\right|\left.-,0\right>=1~~~~,~~~~\left<+,0\right|\left.-,0\right>=0~.\nonumber
\eea
for the mass eigenstates, and 
\bea
\left|a\right>&=&\cos\alpha\left|+\right>-\sin\alpha\left|-\right>\\
\left|b\right>&=&\sin\alpha\left|+\right>+\cos\alpha\left|-\right>~,\nonumber
\eea
for the flavour eigenstates.
For the field configurations (\ref{initial}), the phases are $\varphi_\pm=\vec k\cdot\vec r-\omega_\pm t$ and it is easy to check that the flavour oscillation probability 
has the known expression
\be\label{proba}
P_{a\to b}^{plane}(t)\equiv\left|\left<b,0\right|\left.a\right>\right|^2=\sin^2(2\alpha)\sin^2\left(\frac{1}{2}(\omega_+-\omega_-)t\right)~,
\ee
and is independent of the spatial coordinates.
If one considers a beam of ultra-relativistic particles, the propagation time can be identified with the propagation length, in a system of units in which $c=1$.
The oscillation length is then
\be\label{length}
L=\frac{2\pi}{\omega_+-\omega_-}\simeq\frac{4\pi k}{\Delta m^2}~,
\ee
and corresponds to the experimental observable.

\subsection{Gravity-induced phase shift}

In the presence of gravity, the phase of mass eigenstates depends on the spatial coordinates in a non-trivial way. We assume then that the mass eigenstates have the same amplitude, which is a valid approximation if one considers the situation where the masses are almost-degenerate \cite{Beuthe:2001rc}. 
For the stationary scattering process we consider, the time dependence of mass eigenstates is trivial, and a configuration can be written
\be
\phi_\pm=e^{i\Phi_\pm(\vec r)-i\omega_\pm t}A_\pm(\vec r)~,
\ee
where $\vec r$ denotes the set of space coordinates.
Assuming $A_+\simeq A_-$, the amplitude can be factorised in the linear combinations representing the flavour eigenstates. This amplitude therefore cancels out in the normalisation of states for the calculation of oscillation probability, which thus depends on the phases $\Phi_\pm-\omega_\pm t$ only.
As a consequence, the same argument as the one leading to eq.(\ref{proba}) gives here
\be\label{probagravity}
P_{a\to b}=\sin^2(2\alpha)\sin^2\left(\frac{1}{2}(\omega_+-\omega_-)t-\frac{1}{2}\Delta(\vec r)\right)~,
\ee
where the phase shift we are interested in is
\be\label{shift}
\Delta(\vec r)=\Phi_+(\vec r)-\Phi_-(\vec r)~,
\ee
and is the main subject of our numerical analysis presented in section \ref{sec-numrel}. Time cannot be replaced by the distance travelled in
the probability (\ref{probagravity}), since one doesn't consider specific geodesics, but one can nevertheless characterise the motion of surfaces of constant probability (\ref{probagravity}). 
For this, at a given time $t$ one denotes by ${\cal D}$ a surface corresponding to a given probability, far enough from the BH for spacetime to be assumed flat.
As time increases $\Delta$ must compensate the change in $(\omega_+-\omega_-)t$, for the probability to stay constant on ${\cal D}$, which must therefore be shifted and deformed in such a way that 
\be\label{domainvelocity}
\frac{dP_{a\to b}}{dt}=0~~\Longrightarrow~~\omega_+-\omega_-=\vec v\cdot\vec\nabla\Delta~,
\ee
where $\vec v$ is the velocity of points on ${\cal D}$. The gradient $\vec\nabla\Delta$ is perpendicular to ${\cal D}$ at each point, and its scalar product with the velocity of ${\cal D}$ at this point is therefore a constant characterising the system. For a stationary process as we consider here, the surfaces ${\cal D}$ are periodically generated and move away from the scattering centre.

\subsection{Neutrino interferometry in Schwarzschild geometry}
\label{sec:NIinSchwarz}

Before turning to the study of a massive wave scattered by a BH, we describe here the point of view of particles traveling along specific trajectories.
The first calculation on neutrino interferometry was done in \cite{Fornengo:1996ef}, and related comments can be found in 
\cite{Crocker:2003cw}, \cite{Godunov:2009ce}, \cite{Linet:2002wp} and \cite{Zhang:2000ik}.

We consider the motion of a test particle of mass $m$ in the Schwarzschild metric
\be\label{Schwarzschild}
ds^2=g_{\mu\nu}dx^\mu dx^\nu=f(r)dt^2-\frac{dr^2}{f(r)}-r^2d\theta^2-r^2\sin^2\theta ~d\varphi^2~,
\ee
with
\be
f(r)=1-\frac{R_s}{r}~,
\ee
where $R_s = 2GM$ and we restrict the study to the equatorial plane $\theta=\pi/2$, where the conserved quantities are 
\be
p_t=mf(r)\frac{dt}{ds}~~~~\mbox{and}~~~~p_\varphi=-m r^2\frac{d\varphi}{ds}~.
\ee
Consider a neutrino source at the position $\vec r_A$ where the beam is split into two beams, which are later lensed by a BH, to meet again at the position $\vec r_B$.
The flavour oscillation probability at the intersection point has been calculated in \cite{Fornengo:1996ef}, after evaluating the phase 
along null geodesics. Assuming that the impact parameters $b_1,b_2$ of the two intersecting beams are large compared to $R_s$, and up to terms of order $(m_\pm/k)^2$, the result is
\bea\label{probacurved}
P_{a\to b}^{BH}&=&\sin^2(2\alpha)\left[\sin^2\left(\frac{\Delta m^2}{4k}\left(R+R_s-\frac{R\sum b^2}{4r_Ar_B}\right)\right)
\cos\left(\frac{m_+^2R\Delta b^2}{8k r_Ar_B}\right)\cos\left(\frac{m_-^2R\Delta b^2}{8k r_Ar_B}\right)\right.\nn
&&~~~~~~~~~~~~~~~~~~\left.+\sin^2\left(\frac{\sum m^2R\Delta b^2}{16k r_Ar_B}\right)\sin^2\left(\frac{\Delta m^2R\Delta b^2}{16k r_Ar_B}\right)\right]~,
\eea
where $R\equiv r_A+r_B$ and  
\bea
\sum b^2=b_1^2+b_2^2~~&,&~~\Delta b^2=b_1^2-b_2^2\\
\sum m^2=m_+^2+m_-^2~~&,&~~\Delta m^2=m_+^2-m_-^2~.\nonumber
\eea

\begin{figure}[ht]
    \centering
    \includegraphics[width=0.75\textwidth]{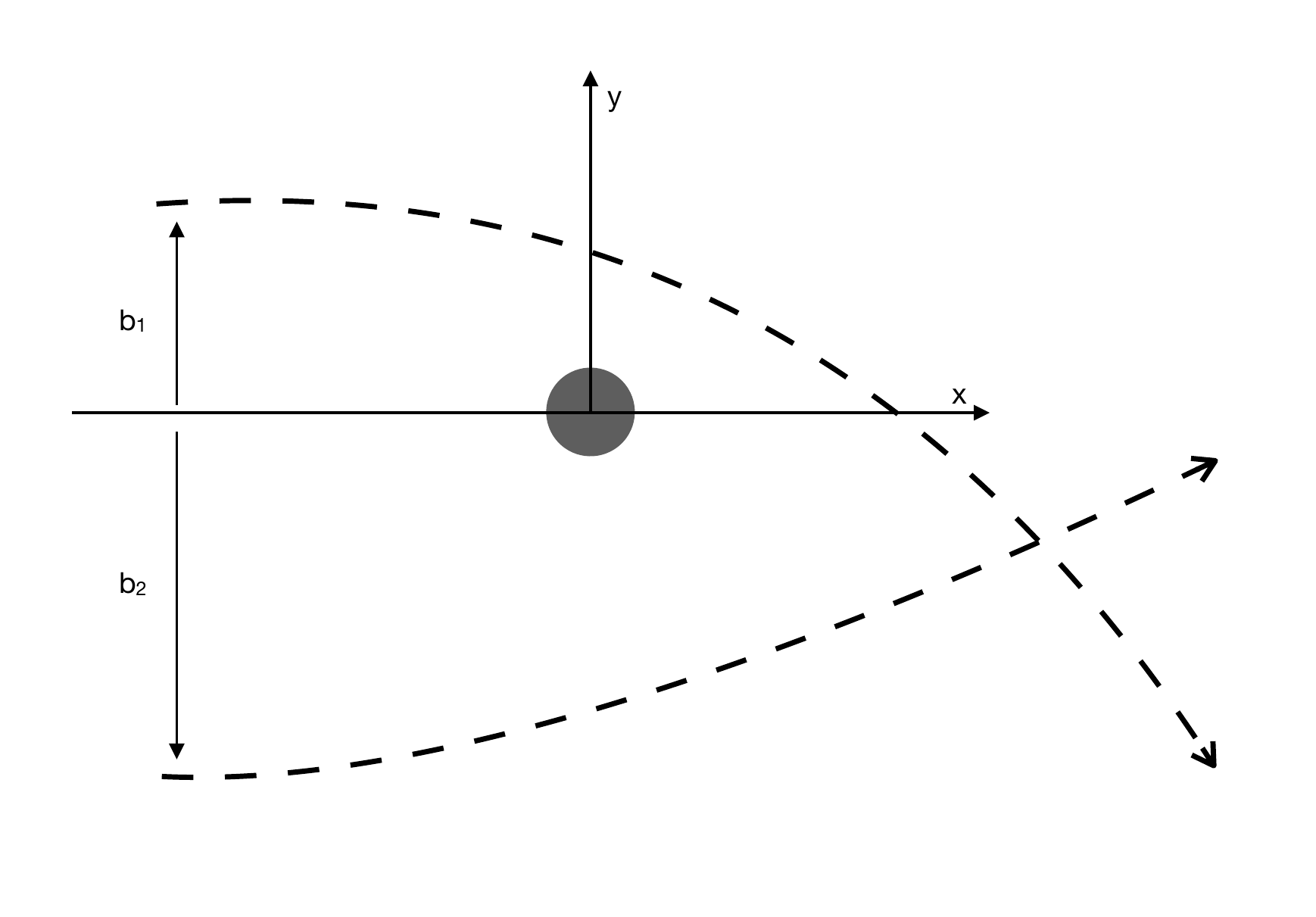}
    \caption{Figure to show the setup described in section \ref{sec:NIinSchwarz}, with initially asymptotically parallel and coherent beams with impact parameters $b_1$ and $b_2$, which are scattered by the BH and intersect for $x \gg R_s$.}
     \label{fig:SketchBH}
\end{figure}

Motivated by a plane wave scattered by a BH, we consider here a source at minus infinity in the x-direction, which produces asymptotically parallel and coherent beams with different impact parameters $-\infty<b<+\infty$.
Far enough from the BH, we can use the asymptotically Cartesian coordinates $(x,y)$ in the equatorial plane, centred on the BH, with the impact parameters 
measured along the $y$ axis, as shown in Fig. \ref{fig:SketchBH}.

The limit $r_A\to\infty$ in the probability (\ref{probacurved}) can be taken in the following way
\be
\frac{\Delta m^2}{4k}(r_A+R_s)=2q\pi~,~~~~q~\mbox{integer and}~q\to\infty~,
\ee
which leads to the oscillation probability at the intersection of two deflected beams 
with impact parameters $b_1,b_2$, given by
\bea\label{probainf}
P_\infty&=&\sin^2(2\alpha)\left[\sin^2\left(\frac{\Delta m^2}{4k}\left(r-\frac{\sum b^2}{4r}\right)\right)
\cos\left(\frac{m_+^2\Delta b^2}{8k r}\right)\cos\left(\frac{m_-^2\Delta b^2}{8k r}\right)\right.\nn
&&~~~~~~~~~~~~~~~~~~~\left.+\sin^2\left(\frac{\sum m^2\Delta b^2}{16k r} \right)\sin^2\left(\frac{\Delta m^2\Delta b^2}{16k r}\right)\right]~,
\eea
where $r$ is the radial coordinate of this intersection point.

\begin{figure}
    \centering
    \begin{subfigure}[b]{0.45\textwidth}
        \includegraphics[width=\textwidth]{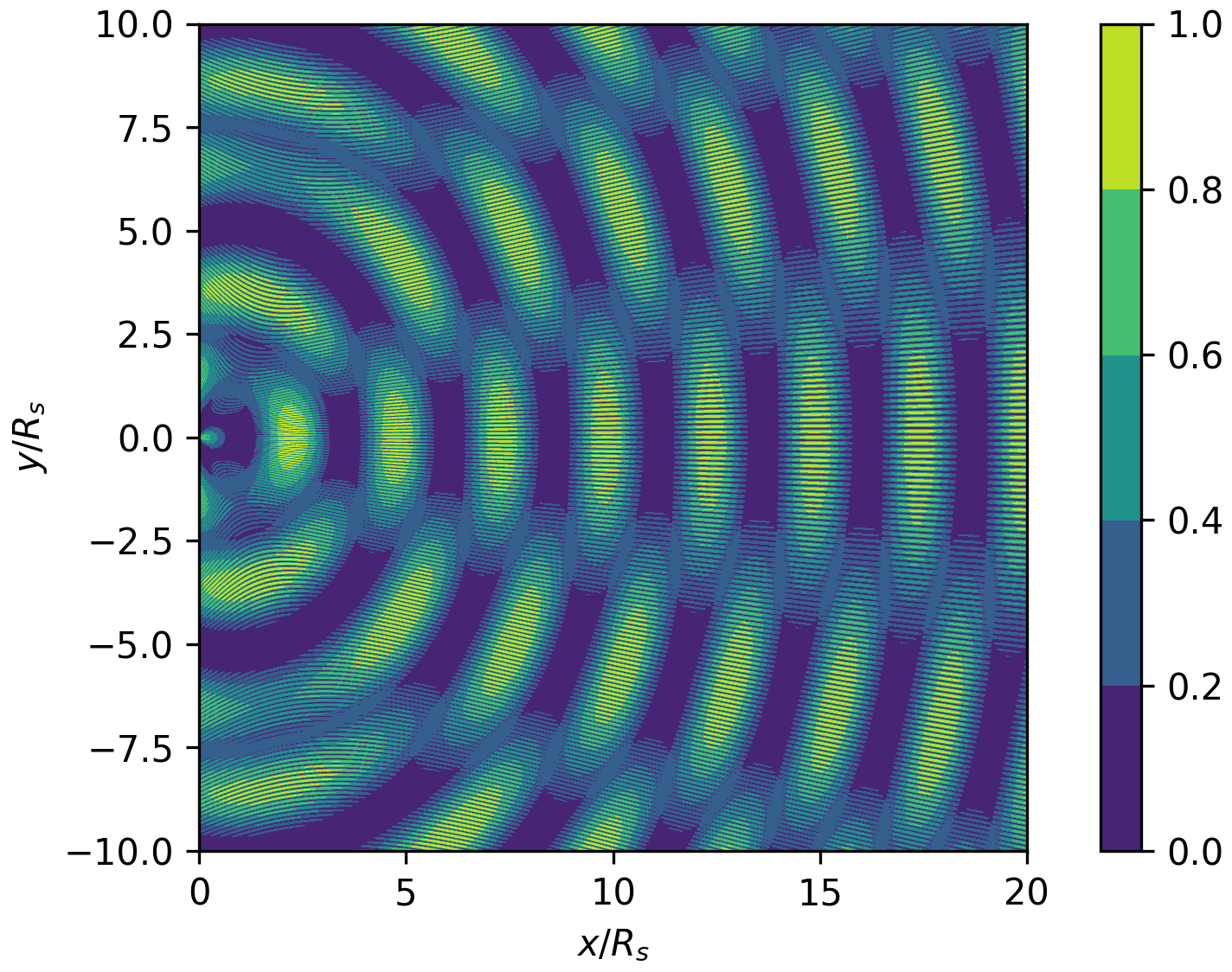}
        \caption{Oscillation length $L \sim 4 R_s$}
    \end{subfigure}
    \begin{subfigure}[b]{0.45\textwidth}
        \includegraphics[width=\textwidth]{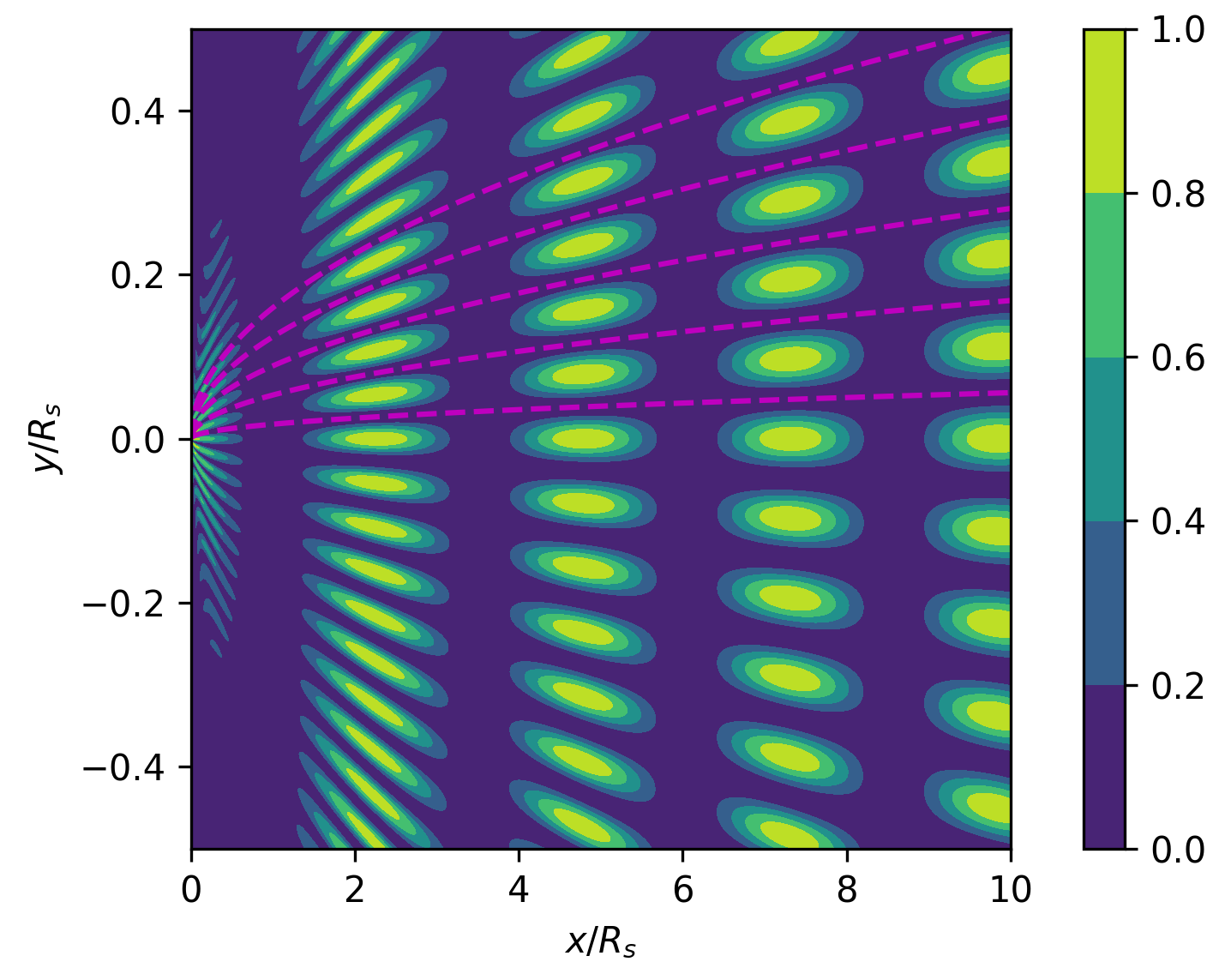}
        \caption{Oscillation length $L \sim 4 R_s$ - magnified}
    \end{subfigure}
    \begin{subfigure}[b]{0.45\textwidth}
        \includegraphics[width=\textwidth]{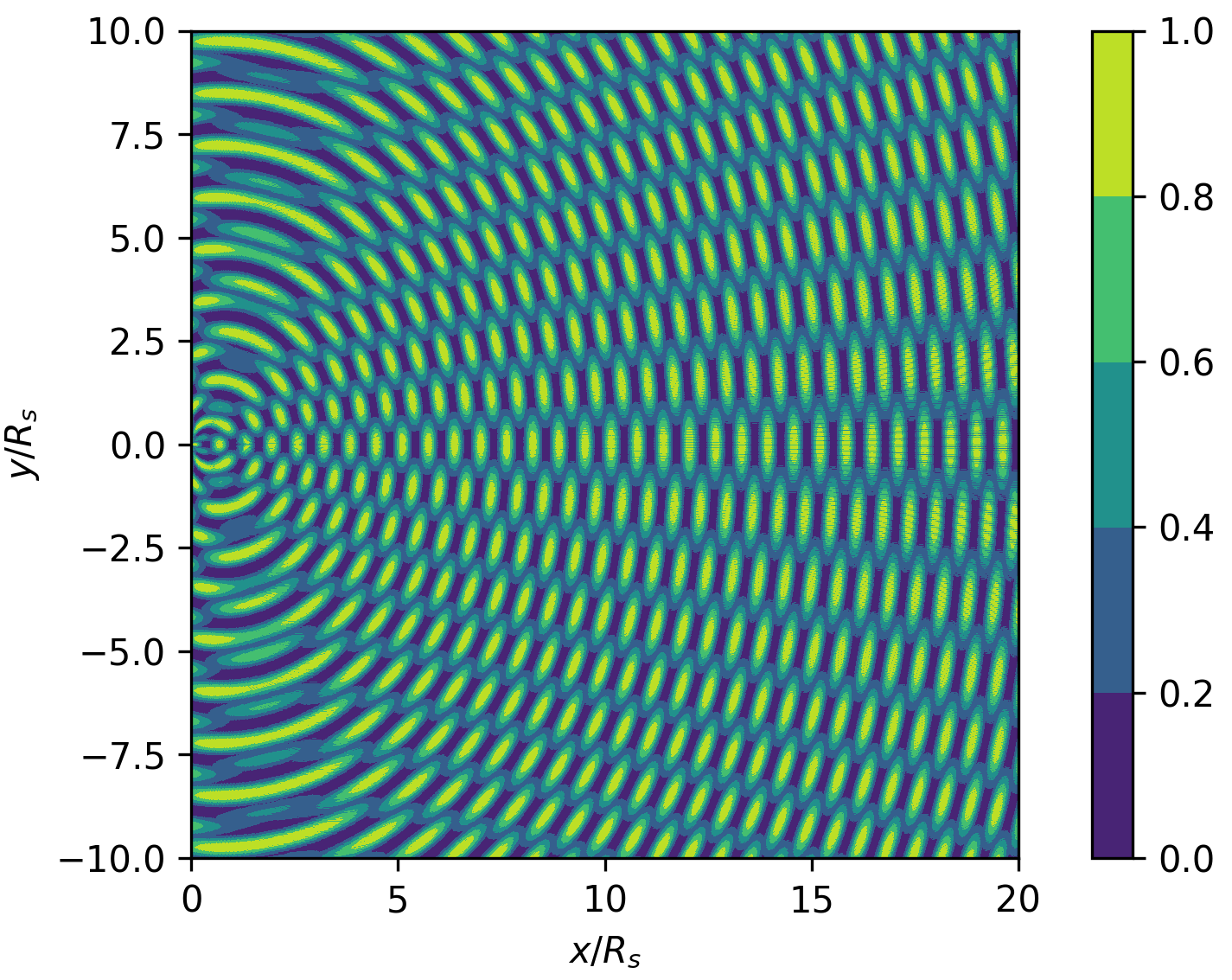}
        \caption{Oscillation length $L \sim R_s$}
    \end{subfigure}
    \begin{subfigure}[b]{0.45\textwidth}
        \includegraphics[width=\textwidth]{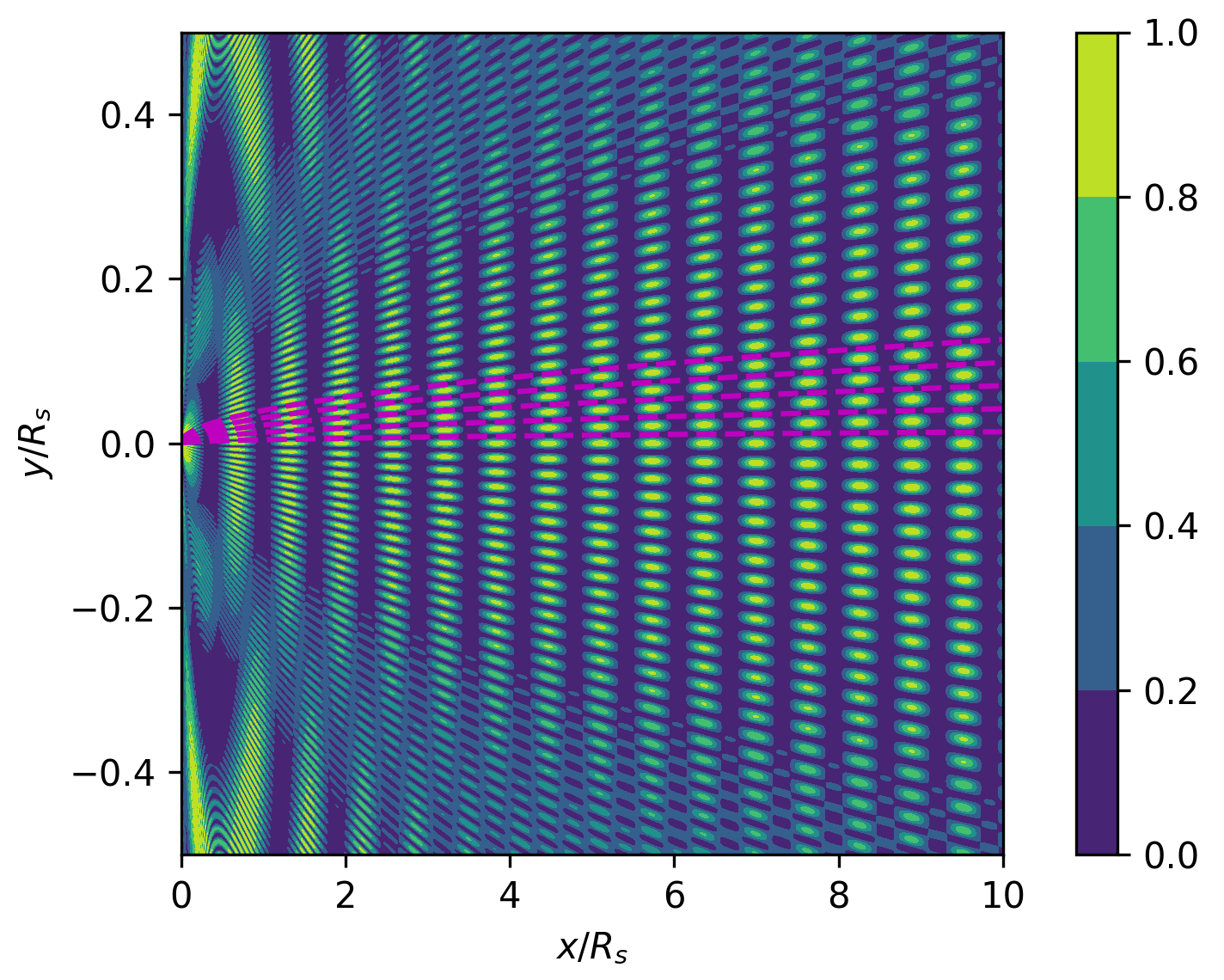}
        \caption{Oscillation length $L \sim R_s$ - magnified}
    \end{subfigure}
    \caption{Plot of the oscillation probability $P_\infty$ given by eq. (\ref{probainf}).
    The BH is located at $x=y=0$ and the horizontal axis $x$ corresponds to 
    the direction of the original momentum of the plane wave. For Figs. (a) and (b), the numerical values used are such that 
    the particles' Compton wavelength is $10^{-10}$ times smaller than the Schwarzschild radius and their initial momentum $k$ is $10^7$ times larger than
    their masses, which differ by 1\%. Fig. (b) - which is stretched along the vertical $y$-axis - shows a magnified section of (a), showing as dotted lines the parabolas of vanishing probability across the first few spherical 
    patterns. The parabolas correspond to the integer values between $n=0$ and $n=4$ in eq. (\ref{parabolas}).
    In figs. (c) and (d) the momentum is 4 times smaller than for figs. (a) and (b), with identical masses, such that the oscillation length in the radial direction is approximately 4 times shorter, and of the order of the BH radius. In this case the parabolas are much closer together. Both Figs. (a) and (c) show a second oscillation feature in the orthoradial direction, where the regions of maximum probability are shifted out of phase.}
     \label{fig:far}
\end{figure}

It is interesting to consider the lines of constant probability in the equatorial plane, assuming that $|b_i|>>R_s$.
The deflection angle $\delta$ for massless particles with impact parameter $b$ along the $y$ axis is
\be\label{deflection}
\delta\simeq-\frac{2R_s}{b}~.
\ee
The equation for the asymptotically straight 
trajectory, after deflection, is then 
\be
\frac{y-b}{x}\simeq-\frac{2R_s}{b}~,
\ee
such that the two impact parameters $b_1,b_2$ corresponding to beams intersecting at the point $(x>0,y)$ are
\bea\label{b1b2}
b_1&\simeq&\frac{y}{2}+\frac{1}{2}\sqrt{y^2+8 R_s x}~>0\\
b_2&\simeq&\frac{y}{2}-\frac{1}{2}\sqrt{y^2+8 R_s x}~<0\nonumber~.
\eea
The probability in eq. (\ref{probainf}) can then be sketched, as in Fig.\ref{fig:far}, and
shows a non-trivial interference pattern for $x>0$. Note that only forward scattering is taken into account.
As can be seen on Fig.\ref{fig:far}(a) (and Fig.\ref{fig:far}(c) for smaller initial momentum), the probability is described by a spherical pattern, with a radial oscillation length consistent with the expression (\ref{length}), and which arises from the sine squared in the first line of eq.(\ref{probainf}).
However, one can see in the magnified and stretched views (b) and (d) that there is another oscillation in the orthoradial direction where the probability vanishes along lines which originate from the BH, and extend in almost radial directions.
These lines are described by the vanishing of the product of cosines in the second line of eq.(\ref{probainf}), which happens when
\be\label{condition}
\frac{m^2\Delta b^2}{8kr}=\frac{\pi}{2}+n\pi~~~~~~(n~\mbox{integer})~,
\ee
where $m\simeq m_+\simeq m_-$.
We focus on the lines which are close to the axis of symmetry, therefore where $x>>y$. Together with the expressions (\ref{b1b2}), the condition (\ref{condition}) 
gives the parabolas
\be\label{parabolas}
 \frac{x}{R_s}\simeq A_n\left(\frac{y}{R_s}\right)^2~,
\ee
where
\be
A_n\equiv a_n\left(1+\sqrt{1+1/(4a_n)}\right)~,
\ee
with
\be
a_n\equiv\frac{R_s^2 m^4}{4\pi^2 k^2(1+2n)^2}.
\ee
We note that, since the present derivations are valid close to the axis of symmetry, we have $\Delta m^2\Delta b^2<<16kr$ and thus the 2nd line of eq.(\ref{probainf}) is negligible.

The parabolas (\ref{parabolas}) corresponding to different integers $n$ can be seen on the magnified views in Figs.\ref{fig:far}(b) and (d). Given the cylindrical symmetry of the problem, the BH therefore generates a family of paraboloids of vanishing oscillation probability, along the direction of the original momentum. This additional oscillation length in the orthoradial direction is much smaller than the radial one, since it depends on the masses and not the mass difference.

As can be seen in the non magnified views in Fig. \ref{fig:far}(a) and (c), there is a third oscillation feature, again in the orthoradial direction, where the regions of maximum probability are ''shifted" out of phase. These out of phase maxima arise from the product of sines squared in the 2nd line of eq.(\ref{probainf}), which is neglected to derive the parabolas (\ref{parabolas}). This term becomes of the same order as the 1st line of eq.(\ref{probainf}) in these sectors. The typical oscillation length of these features depends on the mass difference and thus is of the same order as the radial oscillation length.

As a consequence of these new oscillation features, the flavour detected depends not only on the radial distance to the BH, but also on the distance between the observer and the axis of symmetry of the problem.

Finally, note that several independent length scales appear in the problem, and different combinations could lead to similar interference patterns to those shown in Fig.\ref{fig:far}. 
These length scales arise from:
\begin{enumerate}
\item the eigenmasses $m_+\simeq m_-$;
\item the difference $\Delta m^2$;
\item the initial momentum $k$; 
\item the BH mass $M$. 
\end{enumerate}
We therefore have 3 independent dimensionless parameters on which the interference pattern depends, and these are chosen here, for phenomenological purposes, consistently with ultra-relativistic particles with a small mass difference (compared to their masses), and such that the radial oscillation length is of the order of $R_s$. 

\section{Massive wave scattered by a static black hole}\label{massive}
 \label{sec-massivewave}

In this section we describe some analytical features of a massive plane wave scattered by a BH.  
Studies of massless scalar wave scattering from a BH can be found in \cite{Glampedakis:2001cx}, \cite{Batic:2012rm}, \cite{Kanai:2013rga}, \cite{Gussmann:2016mkp}, \cite{Dolan:2017rtj}, and a pedagogical review is given in \cite{Andersson:2000tf}.
As in the situation of light scattering, both forward and backwards glory effects \cite{Matzner:1985rjn} do occur for spin zero particles \cite{Crispino:2009xt}, which is confirmed in this section.

Unlike in the previous section, where the description was done in the equatorial plane $\theta=\pi/2$, we follow convention for these studies and consider here a cylindrically symmetric system where the wave vector of the original plane wave is along the $z$-axis. In coordinates (\ref{Schwarzschild}), the 
fields therefore depends on $r$ and $\theta$, but not $\varphi$.

\subsection{Decomposition in partial modes}

We consider here a mass eigenstate $\phi$ with mass $m$, satisfying the equation
\be\label{originalevol}
\frac{1}{\sqrt g}\partial_\mu(\sqrt g g^{\mu\nu}\partial_\nu\phi)+m^2\phi=0~,
\ee
where the metric is given by eq.(\ref{Schwarzschild}).
The so-called tortoise coordinate $r_\star$ can then be introduced as
\be
\frac{dr}{dr_\star}\equiv f(r)~,
\ee
with which the metric is conformally equivalent to
\be
d\tilde s^2=(dt+dr_\star)(dt-dr_\star)-\frac{r^2}{f(r)}(d\theta^2+\sin^2\theta~ d\varphi^2)~,
\ee
and involves the Eddington-Finkelstein coordinates $t\pm r_\star$.
In terms of the radial coordinate $r_\star$, the wave equation for a cylindrically symmetric field reads
\be\label{evolbis}
\partial_0^2\hat\phi-\partial_{r_\star}^2\hat\phi+f(r)\left(m^2+\frac{R_s}{r^3}\right)\hat\phi
-\frac{f(r)}{r^2\sin\theta}\partial_\theta(\sin\theta\partial_\theta\hat\phi)=0~,
\ee
where $\hat\phi\equiv r\phi$ does not depend on $\varphi$. The next step is to 
decompose the cylindrically symmetric field $\hat\phi$ on the Legendre polynomials basis
\be\label{expandP}
\hat\phi=e^{-i\omega t}\sum_{l=0}^\infty(2l+1)P_l(\cos\theta) u_l(r)~,
\ee
where the trivial time dependence describes a stationary process.
Given the identity
\be
\frac{d}{dx}\left((1-x^2)\frac{dP_l(x)}{dx}\right)+l(l+1)P_l(x)=0~,
\ee
the expansion (\ref{expandP}) plugged in the evolution equation, eq. (\ref{evolbis}) gives, for every $l$,
\be\label{equaul}
\frac{d^2u_l}{dr_\star^2}+\Big(\omega^2-V_l(r)\Big)u_l=0~,
\ee
where the effective potential seen by each mode $u_l$ is, 
\be\label{effpot}
V_l(r)=f(r)\left(m^2+\frac{l(l+1)}{r^2}+\frac{R_s}{r^3}\right)~.
\ee
Although the coordinate $r_\star$ leads to an elegant formulation of the problem, it is not appropriate for the study at large distance $r>>R_s$ in the massive case, as explained in Appendix A.

\subsection{Semiclassical approach}

In the semiclassical approximation discussed in detail in \cite{FW}, the variable $l$ is replaced by a continuous parameter, with a dominant contribution $\hat l$
obtained from the stationary phase method for the calculation of the phase shift of the mode $u_l$. $\hat l$ satisfies 
\be
\delta(\hat l)=\theta~,
\ee
where $\delta(l)$ is the classical deflection angle and $\theta$ is the observation angle, and the differential cross section of the scattering process is then shown to be
\be
\left.\frac{d\sigma}{d\Omega}\right|_{semiclassical}\simeq\frac{\omega^2(\hat l+1/2)}{\sin\theta(d\delta/dl)_{\hat l}}~.
\ee
If one compares this expression with the classical differential cross section for scattered particles 
\be
\left.\frac{d\sigma}{d\Omega}\right|_{classical}=\frac{|b|}{\sin\theta}\frac{d|b|}{d\theta}~,
\ee
one can make the identification
\be\label{b}
|b|\omega\simeq\hat l+\frac{1}{2}~.
\ee
The latter equation gives an important insight into the scattering process: the dominant mode $u_{\hat l}$ in the scattering process corresponds 
to a beam with typical impact factor given by eq.(\ref{b}).

\subsection{Vicinity of the horizon}

Near the BH horizon, the radial coordinate can be written 
\be
r=R_s(1+\epsilon)~~~~,~~~~\mbox{with}~~~~ 0<\epsilon<<1~,
\ee
and one would expect the potential $(\ref{effpot})$ to be negligible compared to $\omega^2$. One should check though, for fixed $r$, that large values of $l$ do not invalidate this approximation.
Given that $3\sqrt3R_s/2$ is the critical impact parameter for a null geodesic, one can assume that the dominant beams contributing to the scattering pattern in the vicinity of the BH are those corresponding to an impact parameter satisfying $|b|\le3\sqrt3R_s/2$. A detailed discussion of critical impact parameters can be found in \cite{Dolan:2006vj}, which includes the case of massive particles. Since we assume ultra-relativist particles though, the null geodesic approximation is enough for the present discussion.

According to the semiclassical argument above, the dominant modes are then characterised by
\be\label{rangel}
\hat l\le3\sqrt3R_s\omega/2~,
\ee
and satisfy 
\be\label{smalll}
f(r)\frac{\hat l(\hat l+1)}{r^2}\simeq\frac{\epsilon~ \hat l^2}{R_s^2}\le\frac{27}{4}\epsilon~\omega^2<<\omega^2~.
\ee
As a consequence, the solution to eq.(\ref{equaul}) is then approximately independent of $l$ in the relevant range (\ref{rangel}).

If we assume an ingoing wave only (since no signal can escape this region), we obtain $u_l\propto\exp(-i\omega r_\star)$, and the full solution is then approximately
\be
\phi\simeq A(\theta)\exp[-i\omega (t+ r_\star)]~,
\ee
where
\be
A(\theta)=\frac{1}{R_s}\sum_{l=0}^{3\sqrt3R_s\omega/2}(2l+1)P_l(\cos\theta)~.
\ee
We note that \cite{Kuchiev:2003ez} and \cite{Kuchiev:2003fy} discuss the possibility of an outgoing massive wave in this regime, in addition to the ingoing wave, with a different amplitude which takes into account a reflection from the Horizon, related to Hawking radiation. This effect is dominant at energies which are comparable to the Hawking temperature, and since the present study is motivated by high-energy neutrinos, we neglect here the outgoing wave.

Assuming that mass eigenstates have the same amplitude, $A_+(\theta)\simeq A_-(\theta)$, the oscillation probability is given by 
\be\label{rings}
P_{a\to b}^{vicinity}\simeq\sin^2(2\alpha)\sin^2\left(\frac{1}{2}(\omega_+-\omega_-)(t+r_\star)\right)~,
\ee
and depends on the Eddington-Finkelstein coordinate $t+r_\star$ only. This is consistent with the numerical analysis in Section \ref{sec-numrel}, where the Kerr-Schild time coordinate $t'$ satisfies $t'+r=t+r_\star$, such that the probability (\ref{rings}) for fixed $t'$ is almost uniform in the vicinity of the BH, since we have there $(\omega_+-\omega_-)r<<1$ (the oscillation timescale is very long). Note that the expression (\ref{rings}) predicts an infinite number of oscillations, for fixed $t$, as one approaches the horizon, where $r_\star\to\infty$. However, this is an artifact of the choice of observer at infinity, for whom $t$ diverges at the horizon. In terms of $t'$, the number of oscillations is finite.

\subsection{Asymptotic solution}\label{largedistance}

Eq. (\ref{equaul}) can be solved asymptotically, by analogy with the scattering problem in a Coulomb potential. The different steps are explained in 
Appendix A, and the solution can be written in terms of the momenta 
\be
k=\sqrt{\omega^2-m^2}~~~~,~~~~p=(2\omega^2-m^2)/k~,
\ee
and the radial coordinate
\be
\rho\equiv r+\frac{pR_s}{2k}\ln(2kr)~.
\ee
If $A$ is a constant amplitude, the solution can then be expressed in terms of the non-scattered plane wave $\phi^{plane}$, and we find
\be\label{sol3}
\phi\simeq\phi^{plane}+\frac{A}{r}h(\theta)\exp\left(ik\rho-i\omega t\right)~,
\ee
where 
\be
h(\theta)\equiv\sum_{l=0}^\infty(2l+1)P_l(\cos\theta)\left(e^{2i\delta_l}-1\right)~,
\ee
and the complex phase $\delta_l$ is expressed in terms of the momentum $p$, as explained in Appendix A.
Note that the plane wave appearing in eq.(\ref{sol3}) is ``distorted'' in the sense that it involves both radial coordinates, $r$ and $\rho$, as can be seen in Appendix A. 
Eq.(\ref{plane}) in this appendix shows that the expression (\ref{sol3}) is valid up to 
terms of order $1/r^2$, and the unavoidable mixture of radial coordinates $r$ and $\rho$ can be understood as a consequence of the long range of the gravitational interaction.   
Also, the expression (\ref{sol3}) is formal only, since the sum appearing in the definition of $h(\theta)$ is actually divergent \cite{Andersson:2000tf}.
But if one considers a ``thick beam'' with maximum impact parameter $B$, instead of an infinite incident plane wave, then the semiclassical argument given above allows to set a cut off $L$ for $l$, defined by $L\simeq B\omega$.

Finally, the result (\ref{sol3}) is valid for each mass eigenstate, for which one can extract the phase as
\be
\mbox{arg}(\phi)=\arctan\left(\frac{\mbox{Im}(\phi)}{\mbox{Re}(\phi)}\right)~,
\ee
but this is left to next section, where the original equations of motion, eq. (\ref{originalevol}), are solved numerically for each mass eigenstate.

\section{Numerical calculation of the oscillation probability}
\label{sec-numrel}

In this section we describe the results of an evolution of the flavour fields on a fixed Schwarzschild BH background. The purpose is to study the interference patterns produced in the oscillation probability as the plane waves pass the BH. We expect, and indeed find, that a stationary pattern is built up dynamically, radiating from the BH as the evolution progresses.

Full details of the numerical methods and convergence tests are provided in Appendix \ref{appendix:GRChombo}.

\subsection{Initial conditions}

The two mass eigenstates of the flavour fields are set up as plane waves in the $x$-direction, as described by eq.(\ref{eqn:planewave}) with $t=0$, superimposed on a fixed background geometry which describes a Schwarzschild BH in Kerr-Schild coordinates $(t', r, \theta, \varphi)$ (see Appendix \ref{appendix:GRChombo}).

Numerical values are chosen such that $m_-\simeq m_+$, in order to have approximately the same amplitude for $\phi_-$ and $\phi_+$. The values in units set by the BH radius $R_s$ are $m_- = 1.0/R_s$ and $m_+ = 0.99/R_s$. The wavenumber is $k=\pi/R_s$, such that the corresponding wavelength of the mass eigenstates is $\lambda=2 R_s$. The resulting oscillation length in flat space corresponding to these values would be of the order of $L \sim 2000 R_s$. 

These values are not phenomenologically realistic, but are set by numerical constraints. To model neutrinos around solar mass BHs, one should set $m \sim 10^{10} /R_s$ and $k \sim 10^{17} /R_s$ as in the previous sections, but this would mean resolving two very disparate timescales, which is computationally intractable using the current method.

There are thus two possible interpretations of the numerical results. Firstly, assuming a solar mass BH, the mass values chosen would correspond to $m \sim 10^{-10}eV$, so very light neutrino-like particles. Alternatively, for neutrinos themselves, the simulated BH would correspond to primordial BHs of order $10^{-10} M_\odot$, although the momentum $k$ should also be made larger in this case.

Despite these issues, we could expect that some features of the numerical solution are still relevant to the neutrino scattering case. We also show results for a higher momentum case $k=2\pi/R_s$ and observe the differences.

\begin{figure}

    \centering
    \begin{subfigure}[b]{0.4515\textwidth}
        \includegraphics[width=\textwidth]{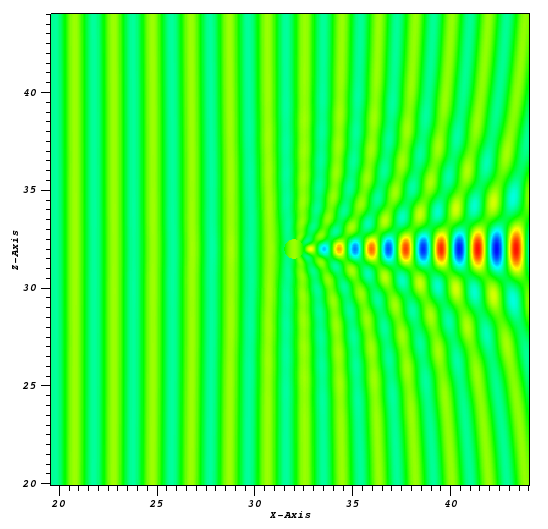}
        \caption{$\phi^+_r$ $k=\pi/R_s$}
    \end{subfigure}
    \centering
    \begin{subfigure}[b]{0.45\textwidth}
        \includegraphics[width=\textwidth]{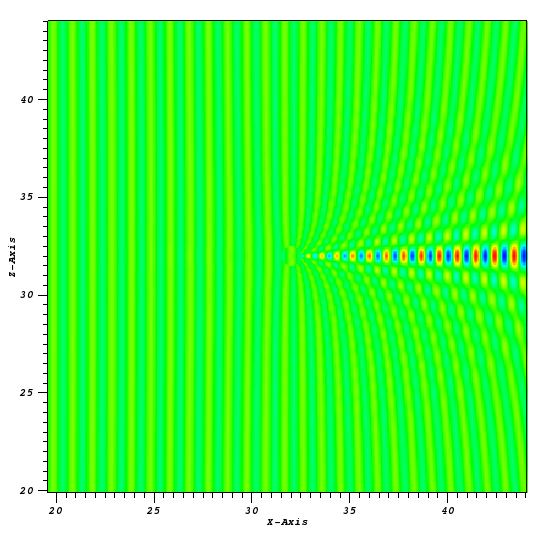}
        \caption{$\phi^+_r$ $k=2\pi/R_s$}
    \end{subfigure}
    \begin{subfigure}[b]{0.4515\textwidth}
        \includegraphics[width=\textwidth]{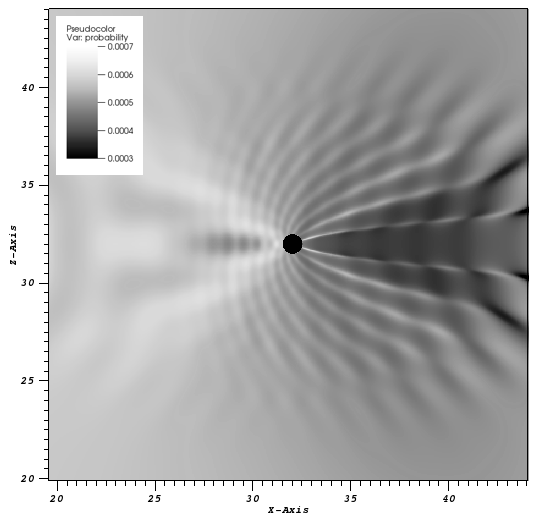}
        \caption{Oscillation probability $k=\pi/R_s$}
    \end{subfigure}
    \centering
    \begin{subfigure}[b]{0.45\textwidth}
        \includegraphics[width=\textwidth]{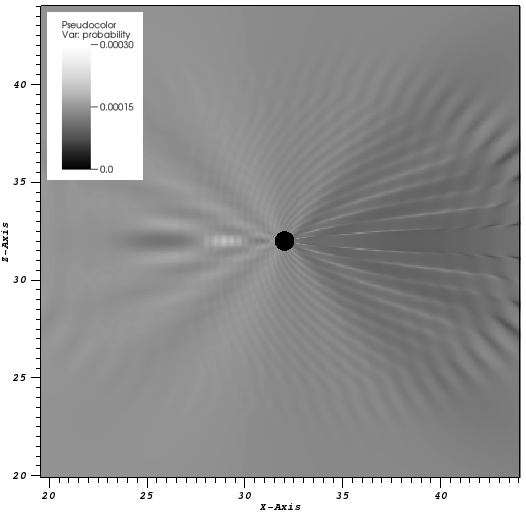}
        \caption{Oscillation probability $k=2\pi/R_s$}
    \end{subfigure}

    \caption{(a) and (c) show plots of the real part of one of the complex fields with coordinates measured in units of the Schwarzschild radius $R_s = 2GM$. The BH is located at $x=z=32$ and the horizontal axis $x$ corresponds to the direction of the original momentum of the plane wave. Plots (c) and (d) show the oscillation probability per eq. (\ref{eqn:numprob}). The scale has been chosen to show the angular pattern which builds up away from the BH, resulting in complex but roughly paraboloidal structures. The plots show waves for which the momenta are $k=\pi/R_s$ and $k=2\pi/R_s$ as indicated. 
    The smoothness of regions at the edges of the plot is an artificial feature, related to the finite run-time of the numerical simulation: scattered waves have not had enough time to escape from the vicinity of the black hole, and thus the interference pattern has not yet built up.}
\label{fig:NumProb}
\end{figure}

\subsection{Evolution and results}

We evolve the mass eigenstates from the initial conditions above according to the Klein Gordon equation eq.(\ref{equamot}), and calculate, at each point of space and for one given time $t'$ the phase difference
\bea
\Theta &=& (\omega_+-\omega_-)t-\Delta(r,\theta)\\
&=& (\omega_+-\omega_-)t'-\tilde\Delta(r,\theta)\nn
&=&\mbox{arg}(\phi_+)-\mbox{arg}(\phi_-) ~.\nonumber
\eea
We thus expect that the system will settle into a stationary regime, with the oscillation probability
\begin{equation}
P_{t'} = \sin^2 {\Theta}~.
\label{eqn:numprob}
\end{equation}
Since the oscillation timescale is of order $2000 R_s$ which is much greater than the simulation time, the term $\tilde\Delta(r,\theta)$ should dominate the probability pattern.

We show the real part of one of the complex fields in Fig. \ref{fig:NumProb} for two different initial momenta $k$. The other fields show a similar behaviour, and we see that the amplitude of the two fields is approximately the same, justifying our use of eq. (\ref{probagravity}) to calculate the probability.

Fig. \ref{fig:NumProb} also shows the spatial patterns of the oscillation probability per eq. (\ref{eqn:numprob}). The results can be summarised as follows:

\begin{itemize}
    \item  The scale in the probability varies by an amount of order $0.001$, thus the region is overall very uniform, as expected from the analytic calculations in Section \ref{sec-massivewave};

    \item In the interference pattern that builds up around the BH, there appear to be two main contributions:

    \begin{enumerate}
        \item Dark lines build up away from the BH, resulting in complex but roughly paraboloidal structures of minimal probability after the BH.
        
        \item Finer, approximately radial lines of maximum probability cross the darker ones. 
    \end{enumerate}
    
    Whilst it is difficult to draw comparisons between the plots here and in Fig. \ref{fig:far}, since the latter uses the particle viewpoint in which time variation is converted to spatial variations along the particle path, one could imagine that the fine structure seen in the magnified plots (on orders much smaller than the oscillation length) is related to the structures seen here numerically. According to eq.(\ref{domainvelocity}), as time evolves (on the order of the oscillation period) paraboloids of constant probability move along the $x$-axis and gradually close up.

    \item The forward glory in the field (the enhanced region in the fields shown in the upper two plots of Fig. \ref{fig:NumProb}), subtends an angle at the BH of $16^o$ in the case of the lower momentum, and this angle halves when the momentum is doubled. This region corresponds to the band of low oscillation probability, between two of the finer radial lines of maximum probability. There is clearly a correspondence between the spatial profile of the forward glory and the oscillation probability, such that the lines of equal probability also become closer together - the structure becomes more fine - with increased momentum. 
    
    \item The amplitude of the forward glory is 1.4 times as large for the higher momentum case, whereas the overall variation in the probability is roughly halved.
    
    \item  Unlike the analytic approximations, the numerical simulation shows the full spatial dependence and includes the backscattering effects of the BH, resulting in a non-trivial pattern of enhanced/suppressed oscillation probability on the left of the BH.

\end{itemize}

\section{Conclusions}

In this work we have modelled the interference patterns in flavour oscillations which are generated as plane waves of neutrinos pass by a BH, using both analytic and numerical techniques.

In all cases we find highly non trivial interference patterns which potentially extend to large distances from the BH. In general, these are paraboloidal shaped regions of enhanced or suppressed probability. These regions, whilst localised, could result in unexpected neutrino detection patterns, in the situation where a BH is roughly in the line of sight between the Earth and an astrophysical event such as a supernova explosion. Together with the information obtained from electromagnetic or gravitational signals, the unexpected neutrino detection pattern could then give essential features on either the BH or the astrophysical event at the origin of the neutrino flux.

The interference patterns we describe assume coupled massive scalars. A full analysis, using fermions, would give the same result in the non spinning case. However, in the case of a spinning BH, one would need to account for the spin flip effect in addition to the flavour oscillation phenomenon. As in flat space time and in the presence of a magnetic field \cite{Schechter:1981hw}, the resulting flavour oscillation probability would be further modulated, due to a new length scale introduced in the system by the BH spin.

Another simplification is that we consider plane waves of coupled flavours, whereas a more phenomenologically motivated model would consider wave packets. This would be more complicated analytically, but potentially simpler to study numerically, since it is localised and thus would not require a high resolution to be maintained across the whole grid. However, the plane wave situation is a first step in understanding the whole interference structure, which would partially be reconstructed by a wave packet passing through a BH.

We also assume a small difference in the eigenmasses of $1\%$, such that we can consider only the phase difference between the mass eigenstates and neglect the amplitude contributions to the probability. We will consider the impact of relaxing this assumption in future work.

Finally, due to computational constraints, the numerical study necessarily focused on the regions close to the BH, with the BH horizon scale corresponding to the initial momenta of the plane waves, rather than the oscillation length as in the analytic work. This is phenomenologically less well motivated, either corresponding to neutrinos around low mass, primordial BHs, or ultra-light neutrino-like particles interacting with solar mass BHs. We hope to develop the numerical methods to study more phenomenologically motivated scenarios in future work. However, the present method of study already provides a means by which to study more complex geometries in the future, in particular spinning BHs, for which analytic derivations are intractable.

\section{Acknowledgements}

JA acknowledges support from the STFC grant ST/ P000258/1. KC acknowledges the computer resources at Marenostrum IV, La Palma and Cierzo and the technical support provided by the Barcelona Supercomputing Center via the PRACE grant Tier-0 PPFPWG, BSC/RES grants AECT-2018-1-0014 and AECT-2017-2-0011. Simulations were also performed on the GWDG cluster in G{\"o}ttingen.

\begin{appendix}

\section{Appendix: Asymptotic solution for a scattered massive wave}\label{appendixA}

Eq. (\ref{equaul}) is usually solved asymptotically, by analogy with the scattering problem in a Coulomb potential. 
This problem involves an equation of the form
\be\label{Coulomb}
\frac{d^2u_l}{dr^2}+\left(\kappa_1^2+\frac{2\kappa_2}{r}-\frac{l(l+1)}{r^2}\right)u_l=0~,
\ee
and is solved in the book \cite{LL} by Landau \& Lifshitz, with asymptotic solutions 
\bea\label{Coulombsol}
u_l&\simeq&\exp\left\{\pm i\left(\kappa_1r+\frac{\kappa_2}{\kappa_1}\ln(2\kappa_1r)-\frac{l\pi}{2}+\eta_l\right)\right\}\\
\eta_l&=&\mbox{arg}~\Gamma[l+1-i\kappa_2/\kappa_1]~.\nonumber
\eea
Unlike the massless case, where one can replace $r$ by $r_\star$ to find an equation of the form (\ref{Coulomb}) - up to higher order terms in $1/r$ -
eq.(\ref{equaul}) involves 
\be
f(r)m^2=f(r_\star)m^2+{\cal O}\left(\frac{\ln r}{r^2}\right)~,
\ee
and the logarithmic term cannot be omitted, compared to $l(l+1)/r^2$. One can come back to the equation in terms of $r$ though, which is
\bea
&&\frac{d^2u_l}{dr^2}+\left[\omega^2-m^2+\frac{R_s}{r}(2\omega^2-m^2)\right.\\
&&~~~~~~~~~~~~~~~\left.+\frac{R_s^2}{r^2}(3\omega^2-m^2)-\frac{l(l+1)}{r^2}\right]u_l={\cal O}(1/r^3)~,\nonumber
\eea
and has the form of eq.(\ref{Coulomb}) only for values of $l$ satisfying 
\be\label{lregime}
l(l+1)>>R_s^2(3\omega^2-m^2)\sim 3(R_s\omega)^2~.
\ee
According to the semiclassical argument though, the latter condition is satisfied for the dominant modes $\hat l$, since the typical impact parameters $b$ 
for beams contributing asymptotically satisfy $b>>R_s$. One can then use the result (\ref{Coulombsol}) to find
\bea\label{assol}
u_l&\simeq&\exp\left\{\pm i\left( kr+\frac{R_s p}{2}\ln(2kr)-\frac{l\pi}{2}+\eta_l\right)\right\}\\
\eta_l&=&\mbox{arg}~\Gamma[l+1-iR_s p/2]\nn
k&=&\sqrt{\omega^2-m^2}\nn
p&=&\frac{2\omega^2-m^2}{\sqrt{\omega^2-m^2}}~,\nonumber
\eea
where $k$ is the momentum at infinity, which is independent of the mass eigenstate, since
\be
k^2=\omega_+^2-m_+^2=\omega_-^2-m_-^2~.
\ee
From eqs.(\ref{expandP}) and (\ref{assol}), the formal expression for a mass eigenstate has the asymptotic form
\be\label{sol1}
\phi\simeq\frac{1}{r}e^{-i\omega t}\sum_{l=0}^\infty(2l+1)P_l(\cos\theta)\left(A_le^{i(k\rho-l\pi/2+\eta_l)}+B_l e^{-i(k\rho-l\pi/2+\eta_l)}\right)~,
\ee
where
\be
\rho\equiv r+R_s\frac{p}{2k}\ln(2kr)~,
\ee
and $A_l,B_l$ are constants of integration. In order to recover usual notations \cite{Andersson:2000tf}, the constants of integration $A_l,B_l$ are traded for 
a global factor $A$ and the complex-valued phase shift $\delta_l$ such that
\be\label{sol2}
\phi\simeq\frac{A}{2ir}e^{-i\omega t}\sum_{l=0}^\infty(2l+1)i^lP_l(\cos\theta)\left(e^{i(k\rho-l\pi/2+2\delta_l)}-e^{-i(k\rho-l\pi/2)}\right)~,
\ee
where 
\be
\mbox{Re}(\delta_l)=\eta_l~.
\ee
The expression (\ref{sol2}) can be compared to the asymptotic form of the plane wave \cite{LL}
\bea\label{plane}
\phi^{plane}&\simeq&\frac{A}{r}e^{-i\omega t}\sum_{l=0}^\infty(2l+1)i^lP_l(\cos\theta)\sin(kr-l\pi/2)\\
&\simeq&\frac{A}{r}e^{-i\omega t}\sum_{l=0}^\infty(2l+1)i^lP_l(\cos\theta)\sin(k\rho-l\pi/2)~,\nonumber
\eea
such that the solution (\ref{sol2}) can finally be written in the form
\be
\phi\simeq\phi^{plane}+\frac{A}{r}h(\theta)e^{i(k\rho-\omega t)}~,
\ee
where
\be
h(\theta)\equiv\sum_{l=0}^\infty(2l+1)P_l(\cos\theta)\left(e^{2i\delta_l}-1\right)~.\nonumber
\ee

\section{Appendix: Numerical Implementation} \label{appendix:GRChombo}

This appendix summarises the key features of the numerical code used in the simulations, which is based on the publicly available numerical relativity (NR) code $\textsc{GRChombo}$, itself built on top of the open source $\mathtt{Chombo}$ framework \cite{Chombo}. For a more full discussion of $\textsc{GRChombo}$ see \cite{Clough:2015sqa}, or the website \url{www.grchombo.org}. Here we describe the key features of the code as used in this work, in which the matter fields were evolved on a fixed BH background, neglecting any backreaction of the fields on the metric (thus note that the NR capabilities of the code were not utilised).

\subsection{Background metric}

We use the Kerr-Schild form of the Schwarzschild solution as the background on which the fields evolve, thus neglecting any backreaction of the flavour fields on the metric. This metric is horizon penetrating and thus simulates the full exterior solution, but necessitates excision at the singularity - in the static Schwarzschild case this can be done by simply setting the flavour fields within the horizon (in practice we do this for $r < R_s/ 2$) to decay to zero. Since the curvature of the metric prevents signals from escaping, errors due to the excision do not (in principle) propagate outwards.

The Kerr-Schild time coordinate $t'$ is related to the Schwarzschild coordinate $t$ by adding the difference between the tortoise coordinate $r_\star$ and the standard Schwarzschild radial coordinate $r$:
\begin{equation}
t' = t + R_S \ln \left(r/R_S - 1\right)
\end{equation}
whilst the radial Kerr Schild coordinate is equal to the Schwarzschild radius (note this is for the non spinning case).

The form of the metric in the standard  $3+1$ ADM decomposition is then:
\begin{equation}
ds^2=-\alpha^2\,dt^2+\gamma_{ij}(dx^i + \beta^i\,dt)(dx^j + \beta^j\,dt)
\end{equation}
where
\begin{equation}
\alpha = (1+2M/r)^{-1/2}
\end{equation}
\begin{equation}
\beta^i = \frac{2 M x^i}{r + 2 M}
\end{equation}
\begin{equation}
\gamma_{ij} = \delta_{ij} + 2 H x_i x_j / r^2
\end{equation}
where $x^i=x_i$ are the cartesian coordinates on the grid and $r^2 = x^2 + y^2 + z^2$.
The trace of the extrinsic curvature, the only other component required for the evolution of the fields, is given by
\begin{equation}
K = 2 \alpha^3 (1+H) x^i \partial_i H + 2 \alpha H \partial_i (x^i/r) ~ .
\end{equation}

\subsection{Matter field evolution}

We have two complex scalar fields, which represent the mass eigenstates of the flavour fields. For each field the real part $\phi_r$ evolves as:
\begin{equation}
\partial_t \phi_{r} = \alpha \Pi_{r} +\beta^i\partial_i \phi_{r} \label{eqn:dtphi2} ~ , \\ 
\end{equation}
\begin{equation}
\partial_t \Pi_{r}=\beta^i\partial_i \Pi_{r} + \alpha\partial_i\partial^i \phi_{r} + \partial_i \phi_{r}\partial^i \alpha +\alpha\left(K\Pi_{r}-\gamma^{ij}\Gamma^k_{ij}\partial_k \phi_{r}+\frac{dV}{d\phi_{r}}\right) \label{eqn:dtphiM2} ~ ,
\end{equation}
where the second order Klein Gordon equation has been decomposed into two first order equations as is usual for numerical evolutions. In the mass eigenstate basis, the mass matrix is diagonal and so the potential gradient is simply given by
\begin{equation}
\frac{dV}{d\phi^\pm_{r}} = m_\pm^2 \phi^\pm_r ~ .
\end{equation}
The imaginary parts follow the same evolution, with $\phi_{r} \rightarrow \phi_{i}$.

\subsection{Numerical details and convergence}
$\textsc{GRChombo}$ is a multi-purpose numerical relativity code, which uses adaptive mesh refinement to increase resolution in areas of interest, and is fully parallelised using the MPI framework. Whilst the full NR capabilities were not required for this work, its flexibility, scaling and built in GR tools made it easy to adapt for the fixed background case.

\subsubsection{Discretisation in space and time}

The metric values and their derivatives are calculated exactly at each point using the analytic expressions above, whereas the flavour field derivatives use 4th order finite difference stencils of the grid values and a 4th order Runge-Kutta time integration for the evolution equations. We use symmetric stencils for spatial derivatives, except for the advection derivatives (of the form $\beta^i \partial_i F$) for which we use one-sided/upwinded stencils. 

The length of the domain is $L=64R_S$, and we use four (2:1) refinement levels with the coarsest having $768^3$ grid points. This coarsest resolution must be sufficient to resolve the wavelengths of the plane wave flavour fields ($\lambda ~ 2R_S$) far from the BH. Around the BH additional resolution is required to properly resolve the horizon area. Kreiss-Oliger dissipation is used to control numerical errors in the evolution of the matter fields, in particular that arising at the grid boundaries.

\subsubsection{Boundary conditions}

We use periodic boundary conditions in all three directions to be able to simulate plane waves travelling in the x-direction. Ideally one would allow the waves to enter and exit at the $\pm x$ boundary by using appropriate ingoing and outgoing boundary conditions, thus minimising the boundary reflections which will inevitably disturb the results. A "lazier" alternative is simply to put the boundaries further away from the region of interest and then only "trust" the region which is causally disconnected from the boundaries at each point in the simulation. Thus the simulation is stopped at $t = L/4$ and only the $r < L/4$ part is considered valid.

For future simulations we plan to develop the appropriate boundary conditions, as this will potentially double the size of the region which we are able to accurately evolve.

\subsubsection{Convergence and testing}

\begin{figure}
    \centering
    \begin{subfigure}[b]{0.45\textwidth}
        \includegraphics[width=\textwidth]{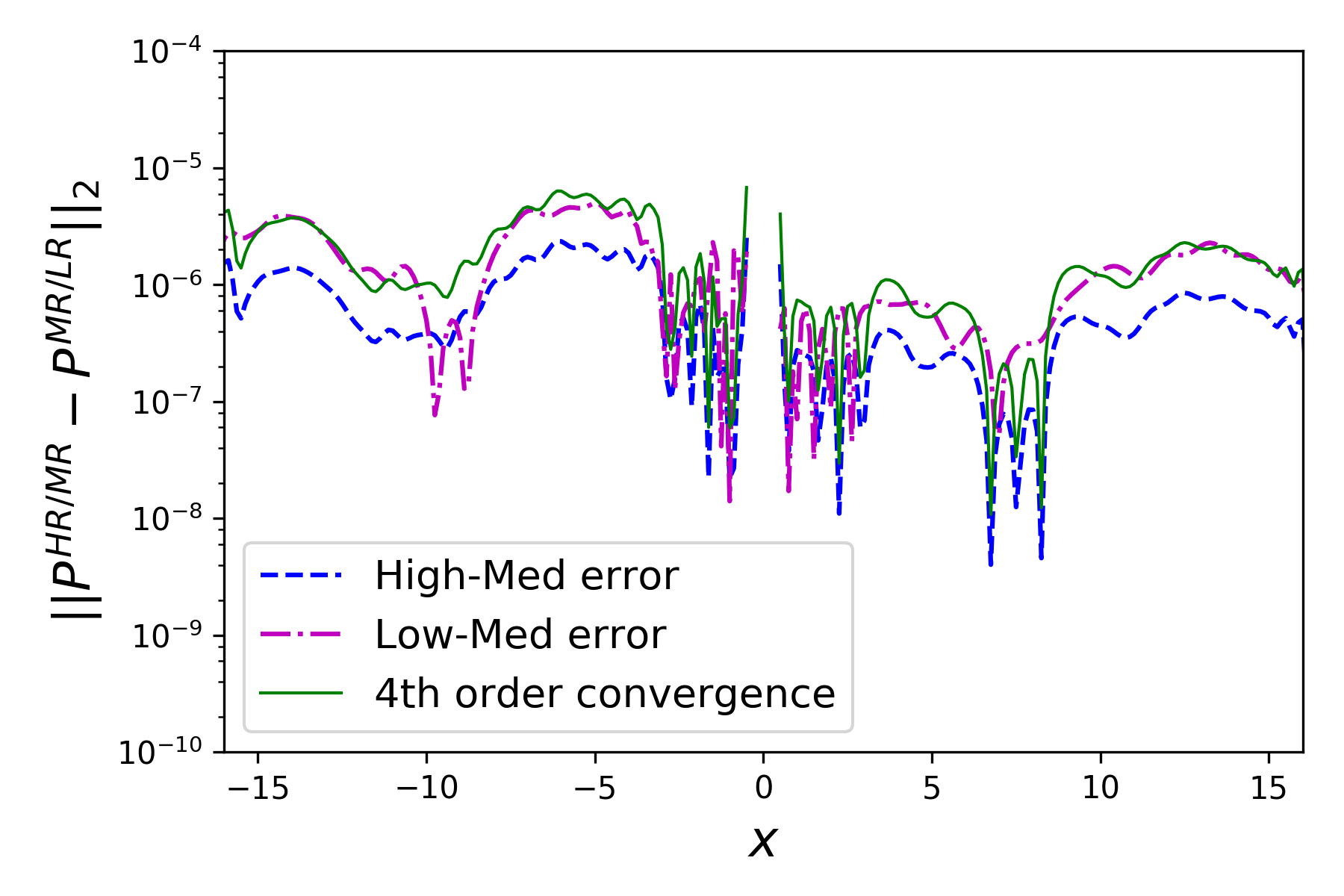}
    \end{subfigure}
    \centering
    \begin{subfigure}[b]{0.45\textwidth}
        \includegraphics[width=\textwidth]{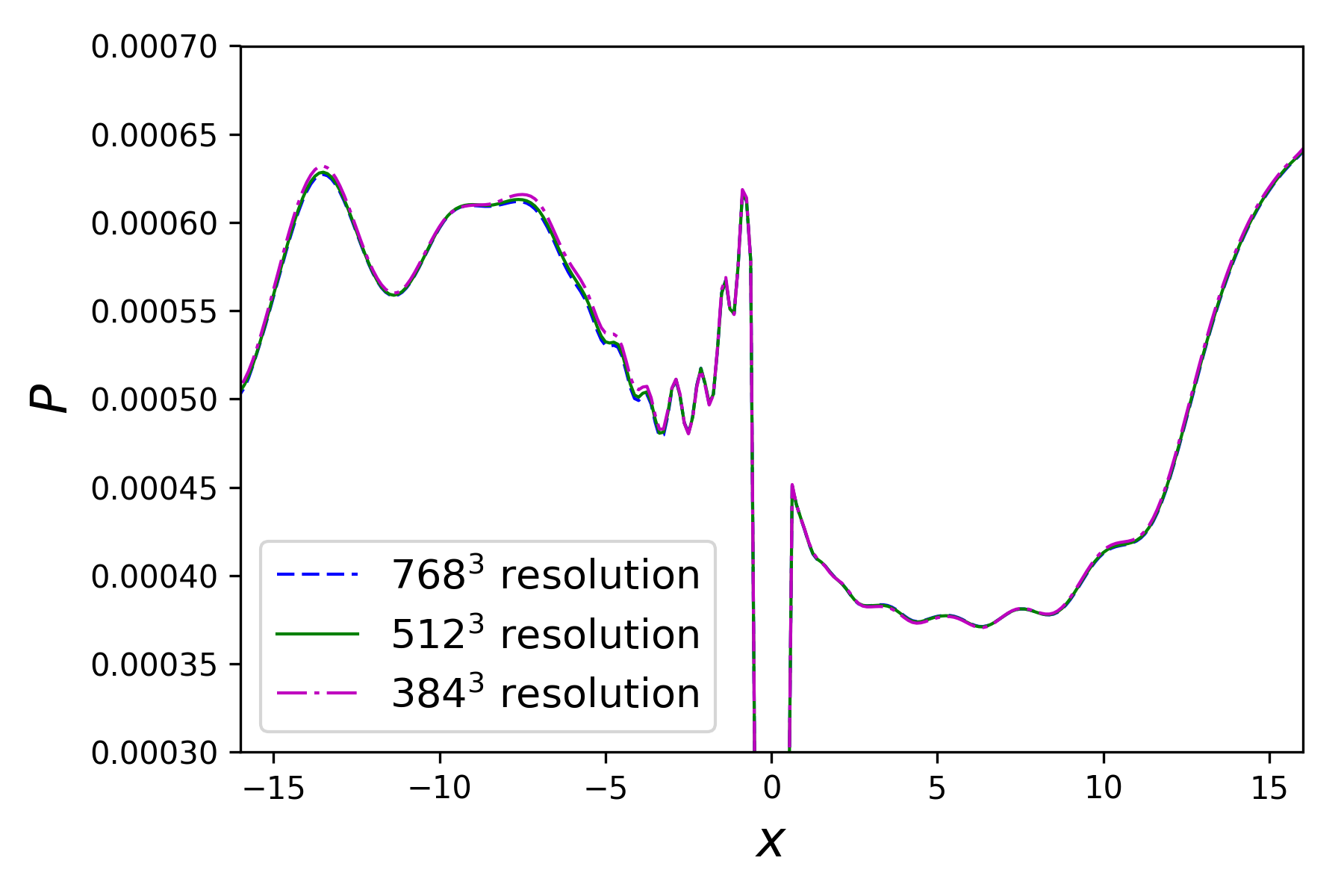}
    \end{subfigure}
    \caption{Here we show that the spatial profile of the probability $P$ is converging at approximately 4th order, by comparing the difference between the results at 3 successive resolutions, which correspond to base resolutions of $384^3$, $512^3$ and $768^3$, each with 4 levels of 2:1 refinement on top.}
\label{fig:Converge}
\end{figure}

We have checked convergence of the simulations, as illustrated in Fig. \ref{fig:Converge}.

We also checked that in the flat space case, the probability remained spatially uniform and the oscillation period was that given by eq. (\ref{proba}) ($\sim 2000 R_s$ for the values used), as expected.

Finally we checked that the solution was unchanged when the boundaries were put twice further out, with $L=128$ and a coarsest resolution of $1024^3$, to confirm that the reflections resulting from the periodicity were not affecting the results within the central area for which we show results.

\end{appendix}

\end{document}